\documentclass[10pt,twocolumn,letterpaper]{article}

\usepackage[pagenumbers]{cvpr} %

\usepackage{xspace}
\newcommand{\shortname}{LATTICE\xspace}
\usepackage{blindtext}
\usepackage{graphicx}
\usepackage{multirow}

\definecolor{cvprblue}{rgb}{0.21,0.49,0.74}
\usepackage[pagebackref,breaklinks,colorlinks,allcolors=cvprblue]{hyperref}

\def\paperID{1855} %
\def\confName{CVPR}
\def\confYear{2026}

\title{LATTICE: Democratize High-Fidelity 3D Generation at Scale}

\author{
    Zeqiang Lai$^{1,2\star}$
    , Yunfei Zhao$^{2\star}$ 
    , Zibo Zhao$^{2}$ 
    , Haolin Liu$^{2}$ \\
     Qingxiang Lin$^{2}$
    , Jingwei Huang$^{2}$ 
    , Chunchao Guo$^{2\ddagger}$ 
    , Xiangyu Yue$^{1\ddagger}$ \\ 
	$^1${MMLab, CUHK} \quad $^2${Tencent Hunyuan}\\
    \url{https://lattice3d.github.io}
}

\begin{document}

\twocolumn[{%
\renewcommand\twocolumn[1][]{#1}%
\maketitle
\begin{center}
    \vspace{-5mm}
    \centering
    \captionsetup{type=figure}
    \includegraphics[width=\linewidth]{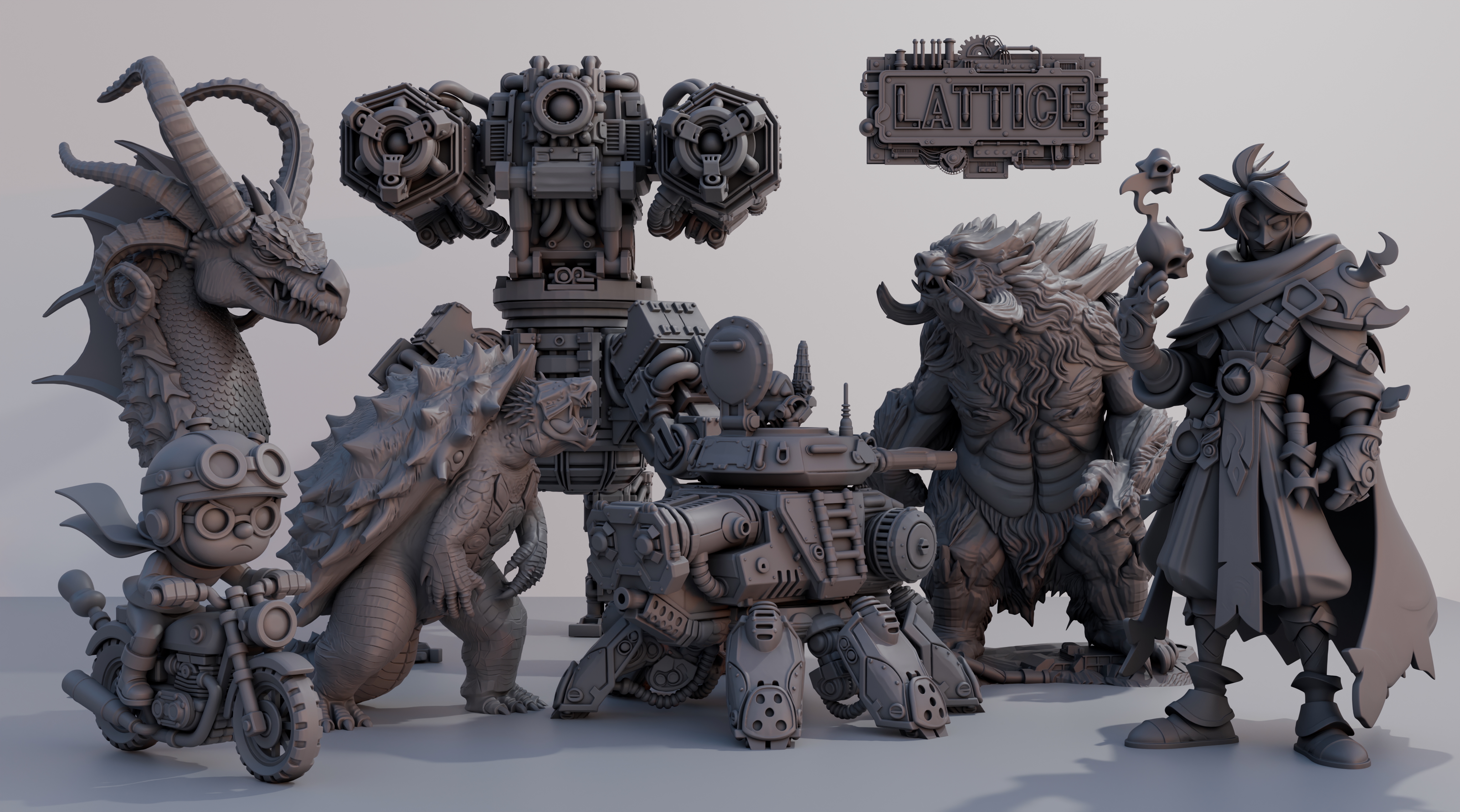}
    \captionof{figure}{
    High quality 3D assets generated by \shortname from a single image.
    }
    \label{fig:teaser}
\end{center}%
}]

\begingroup
\renewcommand\thefootnote{}
\footnotetext{$\star$ Equal contribution. $\dagger$ Corresponding authors.}
\endgroup

\begin{abstract}

We present \shortname, a new framework for high-fidelity 3D asset generation that bridges the quality and scalability gap between 3D and 2D generative models. 
While 2D image synthesis benefits from fixed spatial grids and well-established transformer architectures, 3D generation remains fundamentally more challenging due to the need to predict both spatial structure and detailed geometric surfaces from scratch.
These challenges are exacerbated by the computational complexity of existing 3D representations and the lack of structured and scalable 3D asset encoding schemes. To address this, we propose VoxSet, a semi-structured representation that compresses 3D assets into a compact set of latent vectors anchored to a coarse voxel grid, enabling efficient and position-aware generation. VoxSet retains the simplicity and compression advantages of prior VecSet methods while introducing explicit structure into the latent space, allowing positional embeddings to guide generation and enabling strong token-level test-time scaling. Built upon this representation, \shortname adopts a two-stage pipeline: first generating a sparse voxelized geometry anchor, then producing detailed geometry using a  recitified flow transformer. 
Our method is simple at its core, but supports arbitrary resolution decoding, low-cost training, and flexible inference schemes, achieving state-of-the-art performance on various aspects, and offering a significant step toward scalable, high-quality 3D asset creation.

\end{abstract}

\section{Introduction}
\label{sec:introduction}

Creating high-quality 3D assets is central to modern content pipelines across visual effects, gaming, virtual reality, and industrial design.
Yet, manual creation remains labor-intensive and demands expert skills.
Thus, automating 3D asset generation has become a key challenge at the intersection of vision, graphics, and machine learning.

Despite the impressive progress demonstrated by recent advances in 3D generation~\cite{zhang2024clay, xiang2024structured, zhao2025hunyuan3d}, the question of how to represent 3D assets remains \emph{the “dark cloud” over scalable 3D generation} — a fundamental problem that continues to hinder progress in fidelity, efficiency, and generalization.
This challenge is deeply intertwined \emph{not only with classical 3D representations} — such as meshes, point clouds, Signed Distance Functions (SDFs), radiance fields~\cite{mildenhall2021nerf}, and 3D Gaussian Splattings~\cite{kerbl20233d} — \emph{but also with VAE~\cite{esser2021taming} representations} adopted by latent diffusion models~\cite{rombach2022high}, a key paradigm underpinning the recent advances in 3D generation, as well as in image and video synthesis~\cite{flux2024,kong2024hunyuanvideo}. 
Even with latent diffusion–based compression, the underlying complexity of 3D structure and its cubic growth in memory and computation remain major obstacles, underscoring the need for compact representations.
As a result, \emph{compression, reconstruction, and generation} have become even more crucial in 3D learning, standing as longstanding themes throughout prior research.

The pursuit of such an ideal 3D representation has thus led to what we call {\emph{representation-centric}} research, primarily revolving around compression and reconstruction by two leading paradigms: Sparse Voxel and VecSet.
Sparse Voxel-based methods~\cite{ren2024xcube,xiang2024structured} aim for efficiency by restricting computation to the active voxels near the object surface. However, as reported in Trellis~\cite{xiang2024structured}, even with the inherent sparsity of 3D data, the sequence length of active voxels can be expensively long for training (over 20,000 at $64^3$ resolution), necessitating complex system designs based on sparse convolution~\cite{spconv2022} and attention mechanisms~\cite{liu2021swin, dong2024flex}, which leaves its scalability an open question. Nonetheless, the structured latent space provides strong flexibility for editing and broad generalization to diverse downstream tasks~\cite{engstler2025syncity,yang2025omnipart,li2025voxhammer,huang2025cupid}.
VecSet-based approaches~\cite{zhang20233dshape2vecset, zhang2024clay, zhao2025hunyuan3d} offer a more compact and elegant alternative by compressing 3D objects into a small set of feature vectors via cross-attention between densely sampled point cloud and sparsely sampled point queries -- a set of point coordinates uniformly sampled in the object surface. Remarkably, as few as 3,072 vectors can already yield excellent reconstruction quality, making VecSet-based models highly efficient. 
Moreover, all operations within these models — including the VAE and DiT components — can be implemented using standard self- and cross-attention layers, enabling excellent scalability within modern transformer architectures.
Despite the strong advances in compression and reconstruction, current models remain notably behind 2D latent diffusion models in quality and scalability, leaving the enhancement of 3D generation capability an open and underexplored challenge.

\begin{figure}[t] 
  \centering
\includegraphics[width=\linewidth]{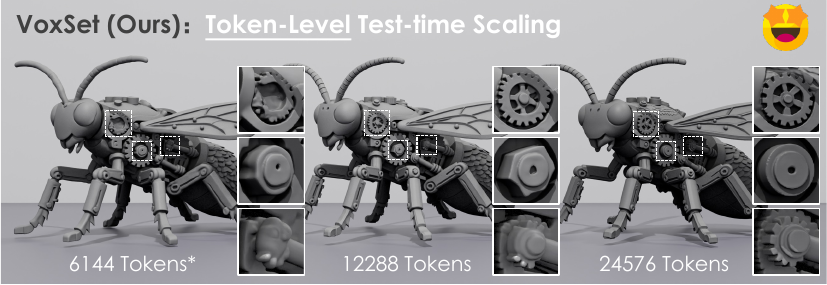}
  \caption{Illustration of test-time scaling in our model. The model is trained with up to 6,144 tokens, but is evaluated under different token counts at test time, showing notable improvements.} 
  \label{fig:scaling2}
\end{figure}

In this paper, we seek to answer a central question in a \emph{generation-centric} perspective, \ie,
\begin{quote}
What truly defines a good representation for 3D diffusion \textbf{generator} itself? 
\end{quote} 
From this viewpoint, we start by asking
--
\textit{why does 3D generation still significantly lag behind 2D in quality and scalability?} At its core, the gap stems from a fundamental difference in how the generative task is framed. 
In 2D image synthesis, the spatial grid is predefined -- models only need to infer RGB values at fixed pixel coordinates (a secret condition that greatly simplifies the denoising processes).
3D generation, however, faces a far more open-ended task: it must discover both \textbf{where} to place content in space and \textbf{what} to represent there (\eg, SDF, RGB).
This joint reasoning over structure and content dramatically expands the search space and introduces ambiguity, making optimization harder and scaling behavior less predictable\footnote{See Appendix for more discussions.}.

At first glance, Sparse Voxel may seem a promising choice due to its inherent spatial structure.
While this holds true, we instead still prefer a \emph{VecSet}-based representation for its distinctive advantages we detailed later, more importantly, its strong capability for \emph{test-time scaling}, as illustrated in Fig.~\ref{fig:scaling2}.
Building upon the insights discussed earlier, we demonstrate that it is possible to combine the best of both worlds through \emph{Localizable Code} -- a unified and high-level abstraction for any representation that tackles the joint reasoning problem.
Crucially, it is localizability rather than structure that truly matters. Guided by this principle, our key idea is to add localizable structure to VecSet~\cite{zhang20233dshape2vecset} latent codes. In other words, we aim to decouple the prediction of \emph{where} and \emph{what}, and guide the unstructured VecSet generation with structure/position, mimicking the success of image generation based on a 2D grid. 

To achieve this, we investigate the positional information secretly encoded in VecSet latent produced by point queries, \ie, each latent is strongly correlated with regions near the position of its corresponding point query, as hinted in \cite{lai2025flashvdm}. However, this information can hardly help during the generation as the positions of point queries are unknown at test time. Therefore, we introduce \emph{voxel queries} as a perfect replacement of point queries. Instead of utilizing point coordinates on the surface, we use the center coordinates of the active voxel intersecting the object's surface. These active voxel grids can be very coarse, thus can be easily obtained, during the test time, by voxelizing an existing geometry generated by any off-the-shelf geometry generation models~\cite{xiang2024structured, hunyuan3d2025hunyuan3d} with less perfect quality.

\begin{figure}[tbp] 
  \centering
  \includegraphics[width=\linewidth]{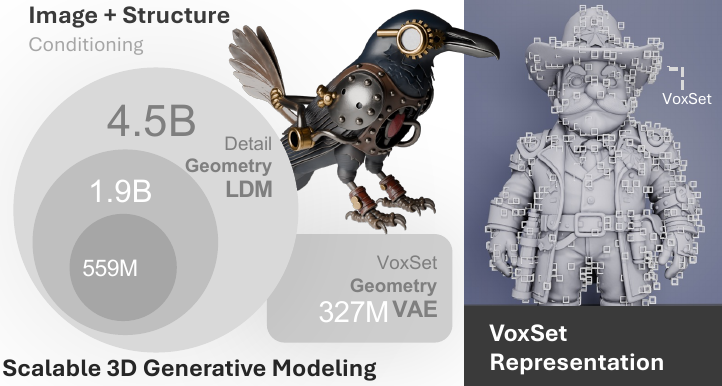}
\caption{\shortname system: At its core is a novel VoxSet representation, enabling scalable 3D modeling from 0.6B to 4.5B.}
  \label{fig:system}
\end{figure}

As a result, our first improvement leads to \emph{VoxSet}, a new semi-structured representation that inherits the efficiency and simplicity of VecSet~\cite{zhang20233dshape2vecset}, while introducing structure into its latent space. This design brings several unique benefits: (1) it enables {\emph{flexible encoding and decoding in abritrary resolution}}, which makes training particularly cheap as multi-stage training is possible by pretraining on very low token size and progressively scaling the tokens up; (2) every latent in {\emph{VoxSet is structured, anchored in a 3D regular voxel grid}} so that the position information can be directly injected into the diffusion transformer (DiT)~\cite{Peebles2022DiT} through positional embedding~\cite{vaswani2017attention,su2024roformer}, which provides strong guidance during diffusion generation and is proven to be essential in model scaling in our experiment.

Based on VoxSet, we present \emph{\shortname}, a general framework designed to generate high-fidelity and detailed 3D assets. \shortname employs a two-stage pipeline. In the first stage, it generates a sparse voxel grid by voxelizing a coarse mesh produced by any off-the-shelf model, such as Hunyuan3D-2~\cite{zhao2025hunyuan3d} or Trellis~\cite{xiang2024structured}. In the second stage, it generates geometry VoxSets at arbitrary resolutions (number of tokens) within the selected voxel grid. Built on rectified flow transformer~\cite{flux2024} and a progressive training strategy, we train a family of large-scale image-to-3D generation models — with up to 4.5 billion parameters, as shown in Fig.\ref{fig:system}, capable of producing detailed meshes from a single image. Through extensive evaluation, we demonstrate that \shortname exhibits strong superiority against previous state-of-the-art models, and is distinguished by several key strengths, as summarized below:
\begin{itemize}
    \item \textbf{Test-time scaling.} Our model exhibits a strong test-time scaling effect. The model trained with up to 6144 tokens/voxel cells can be directly scaled to up to 30720 tokens during the test time, with consistent improvement. 
    \item \textbf{Low-cost training.} Our base model, with 2 billion parameters, can be effectively trained in under 24 hours using 64 GPUs, while still significantly outperforming previous methods.
    \item \textbf{Simplicity.} The model architecture is exceptionally simple — relying solely on a pure Transformer design, without any complex or sparse components.
    \item \textbf{Exceptional performance.} Our model achieves significantly strong performance in 3D generation, excelling in geometry smoothness and detail preservation. 
\end{itemize}

\shortname represents a significant step forward in next-generation 3D assets generation, bridging the gap between generated and handcrafted 3D assets. We hope this work offers valuable insights into effective scaling of 3D generation models and opens up new possibilities for automated, high-fidelity 3D content creation.

\section{Related Works}

\subsection{3D Representations}

Unlike images and videos, which are universally represented by pixel colors, 3D assets exhibit a wide variety of representations tailored to different application contexts. Common atomic representations include voxels, point clouds, Signed Distance Fields (SDF), polygon meshes, DMTet~\cite{shen2021dmtet}, Flexicube~\cite{shen2023flexible}, Neural Radiance Fields (NeRF) \cite{mildenhall2021nerf}, and Gaussian Splatting \cite{kerbl20233d}, among others. These representations, whether explicit or implicit, serve distinct roles in the 3D industry—for example, point clouds are prevalent in perception tasks like autonomous driving~\cite{qi2017pointnet}, NeRF excels in novel view rendering, and polygon meshes remain the standard for gaming and real-time applications. These atomic representations can often be converted between one another—for example, polygon meshes can be extracted from SDFs via the marching cubes algorithm~\cite{lorensen1998marching}. Ultimately, the choice of representation is task-dependent and directly influences network design, \eg, autoregressive models~\cite{weng2024scaling} for meshes, and diffusion models~\cite{zheng2023locally} for SDF. 

Nonetheless, even lightweight and flexible representations such as implicit functions still impose significant modeling and computational burdens on deep neural networks, especially diffusion models~\cite{rombach2022high} -- the current golden paradigm for 3D generation. As a result, the latent representations of 3D assets have emerged as a new research focus, aiming to enhance efficiency. These representations, with difference by their own, can be generally categeoried into three popular types, \ie, (1) VecSet, represented by 3DShape2VecSet~\cite{zhang20233dshape2vecset}, compress 3D shapes into 1D latent sets; (2) Triplane, represented by Direct3D~\cite{hong2023lrm}, compress shapes into three orthogonal features planes; and (3) Sparse Voxel, represented by XCube~\cite{ren2024xcube}, converted 3D assets into features anchored on sparse voxels. 
A long-standing belief holds that the spatial locality of sparse voxel representations helps preserve fine details, whereas VecSet representations, despite their efficiency, tend to lose details due to their global modeling. In this paper, we challenge this idea and identify that the key for 3D generative models lies not in locality, but in a well-known structure at test time. Here, we introduce VoxSet, a semi-structured latent representation that combines efficiency and strong expressiveness.

\subsection{Geometry Generation} 

3D geometry generation has advanced rapidly in recent years. Early works~\citep{wu2016learning,sanghi2022clip,yan2022shapeformer,yin2023shapegpt} based on different generative models~\citep{kingma2013auto,goodfellow2014generative,papamakarios2021normalizing} demonstrated the preliminary potential for generating specific categories of geometry. With the rise of diffusion models~\citep{rombach2022high, ho2020denoising}, 3D geometry generation methods based on score distillation~\citep{poole2022dreamfusion} have been introduced, enabling text-to-3D generation by leveraging text-to-image models. 
Feedforward methods such as LRM~\citep{hong2023lrm}, Hunyuan3D 1.0~\citep{yang2024hunyuan3d}, and LGM~\citep{tang2024lgm} represent another line of research focused on generating 3D assets in a single step. On the other hand, autoregressive models, \eg MeshGPT~\citep{siddiqui2024meshgpt}, BPT~\citep{weng2024scaling}, and Meshtron~\citep{hao2024meshtron} have become popular for mesh generation with human-like topology. 

Recently, native 3D diffusion models have significantly improved generation quality by utilizing 3D data. Notable examples include Michelangelo~\citep{zhao2024michelangelo}, CLAY~\citep{zhang2024clay}, Hunyuan3D 2.0~\cite{zhao2025hunyuan3d}, TripoSG~\cite{li2025triposg}, and Step1X-3D~\cite{li2025step1x}, building on 3DShape2VecSet~\cite{zhang20233dshape2vecset}. Despite great success, these methods seem to struggle at generating highly detailed meshes. On the contray, another line of research, following XCube~\cite{ren2024xcube}, shows promising results by works as Trellis~\citep{xiang2024structured}, Hi3DGen~\cite{ye2025hi3dgen}, SparseFlex~\cite{he2025sparseflex}, Sparc3D~\cite{li2025sparc3d} and Direct3D-s2~\cite{wu2025direct3d}. 
Nevertheless, we show that it is effective scaling through localizable guidance, rather than VecSet or XCube, that matters in detailed geometry generation.

\section{Scalable 3D Generative Modeling}

The goal of \shortname is to explore a new paradigm for scalable 3D generative modeling. To achieve this, a key design choice in our approach is utilizing a coarse geometry structure as strong guidance for detailed geometry generation. This design, despite being used in many voxel-based approaches, such as XCube~\cite{ren2024xcube} and SLAT~\cite{xiang2024structured}, is significantly underestimated. 
In this work, we show that it is actually essential for effective model scaling and performance improvement, no matter what the representation is.

\begin{figure}[t] 
  \centering
  \includegraphics[width=1\linewidth]{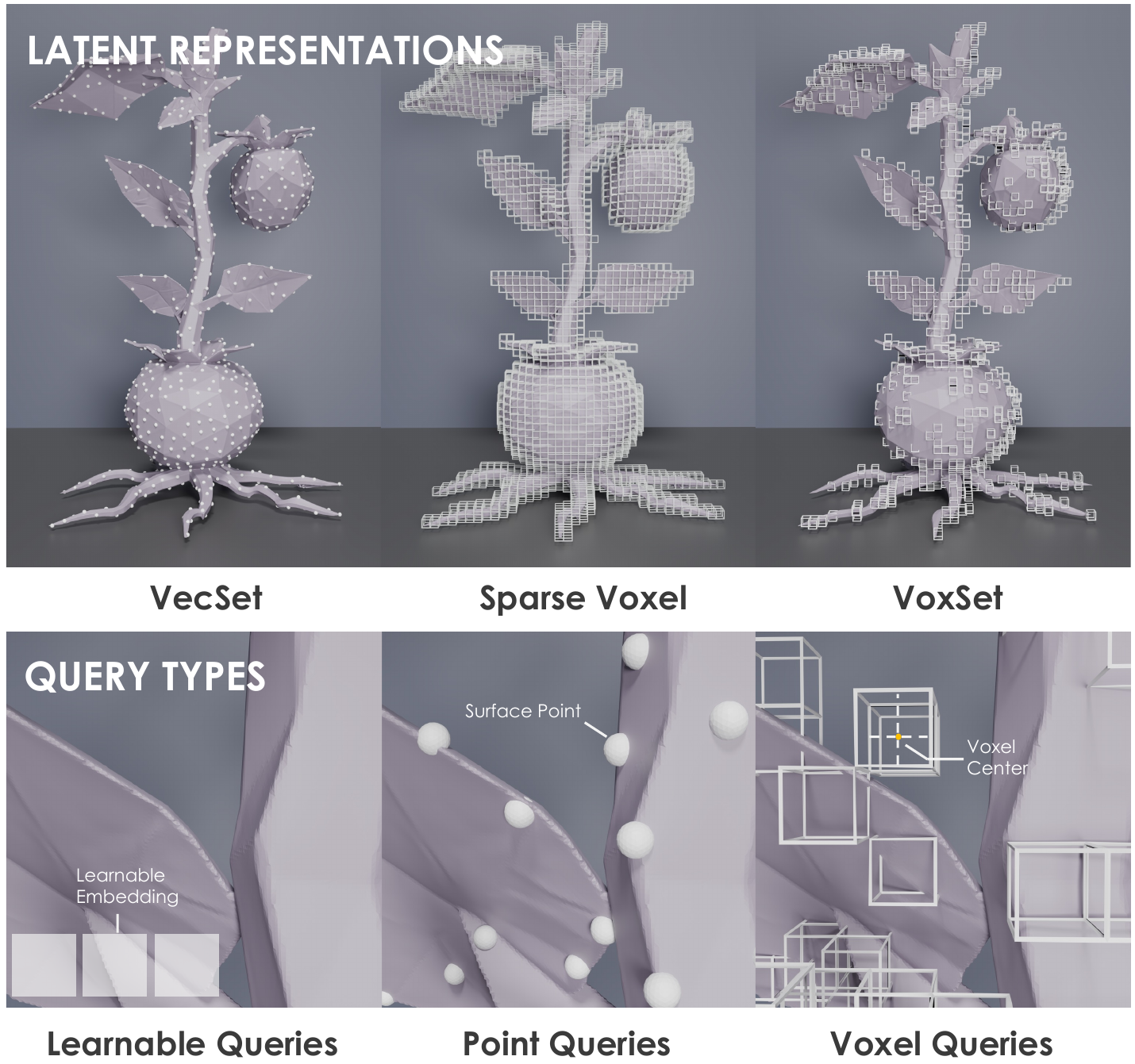}
  \caption{Illustrations of different latent representations and different query types. } 
  \label{fig:voxset}
\end{figure}
\begin{figure*}[htbp] 
  \centering
  \includegraphics[width=\textwidth]{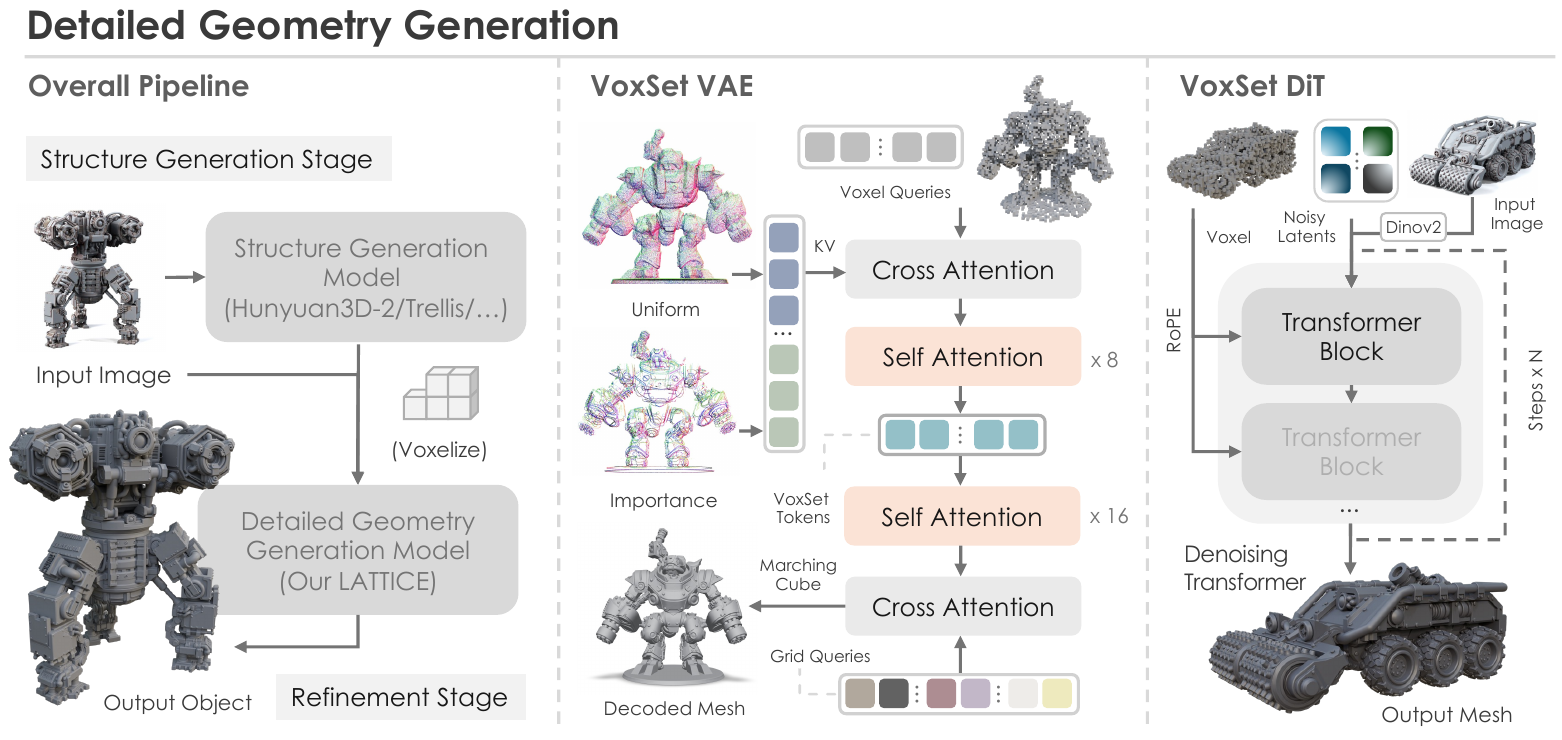}
  \caption{\textbf{\shortname Model Architecture}: it features a two-stage coarse-to-fine pipeline and a novel VoxSet VAE and DiT. } 
  \label{fig:shape_arch}
\end{figure*}

\subsection{VoxSet Representation}

Underlying the architecture of \shortname, it is \emph{VoxSet} representation that builds up the core of the entire system. 
Existing 3D representations—such as meshes, point clouds, signed distance field (SDF), NeRF~\cite{mildenhall2021nerf}, FlexiCubes~\cite{shen2023flexible}, VecSet~\cite{zhang20233dshape2vecset}, and SLAT~\cite{xiang2024structured}—can be broadly categorized into two types. 
The first includes explicit or implicit atomic representations, such as point clouds, SDFs, and NeRF, which directly encode geometry or appearance. The second includes latent representations, such as VecSet and SLAT, which are built upon atomic representations with variational autoencoders (VAE) \cite{kingma2013auto, esser2021taming} and tailored for building compact latent space in latent diffusion models \cite{rombach2022high}.

VoxSet is a latent representation guided by two key principles: scalability and structural latent space.
To support scalability, VoxSet compresses any 3D asset into a sequence of latent tokens via a cross-attention mechanism, following the design of 3DShape2VecSet~\cite{zhang20233dshape2vecset}. 
Formally, given a 3D object, we employ a VAE to encode its point cloud representation and reconstruct the corresponding SDF, from which a surface mesh can be extracted via the Marching Cubes algorithm~\cite{lorensen1998marching}. The input point cloud $P \in \mathbb{R}^{N \times 7}$ captures multiple attributes per point, where $N$ denotes the total number of points and each point encodes its 3D coordinates, surface normal, and a binary sharpness indicator marking whether it lies on a sharp edge. Following the strategy in Hunyuan3D-2~\cite{zhao2025hunyuan3d}, the point cloud is constructed by combining uniform sampling over the surface with importance sampling around sharp edges to better preserve high-frequency details. 

\textbf{Efficient Scaling via Sparsity.} The latent representation, \ie, a token sequence, is obtained via performing cross-attention between the input point cloud and a set of query tokens following a series of self-attention layers. The decoder is designed symmetrically, where the SDF grid coordinates serve as queries of the cross-attention against latent tokens. 
Notably, these latent tokens, in fact, secretly encode the global signals; thus we could represent any 3D object with a latent sequence of any length~\cite{zhang2024clay,chen2024dora,zhao2025hunyuan3d}. This is particularly useful as progressive token scaling (a more fine-grained strategy than progressive resolution scaling) is possible to greatly reduce the training cost, \eg, starting pretraining from 1024 tokens and progressively increasing to more. Even more, as evidenced in FlashVDM~\cite{lai2025flashvdm}, a diffusion model trained on 512 latent tokens can be directly scaled up to 3072 tokens \textbf{in test-time} with better performance, which makes VoxSet particularly economical.

\textbf{Voxel Queries for Detail Modeling.} The choice of query set is a crucial design choice in VecSet-like methods~\cite{zhang20233dshape2vecset,zhao2024michelangelo}, which also serves as a key distinguishing factor of VoxSet. 
In 3DShape2VecSet~\cite{zhang20233dshape2vecset}, two types of query set, \ie, learnable queries and point queries, are proposed. The learnable queries encode the global statistics and are easy to train, but are limited in scaling up for better reconstruction and generation performance. Point queries are downsampled point clouds with furthest point sampling, which encode local information around the queries and support encoding-decoding at arbitrary resolution, favoring low-cost progressive scale-up. 
Moreover, the locality of latent tokens encoded by point queries is very strong and correlated with their position, as discussed in FlashVDM~\cite{lai2025flashvdm}. 

In other words, the latent set from point queries is ordered with position information secretly encoded. 
However, none of existing models~\cite{zhang2024clay,zhao2025hunyuan3d,li2025triposg} utilize the information. 
One of the biggest obstacles is that point queries are sampled on the object surface, whose positions are unknown during test time. 
To address this problem, we introduce \emph{Voxel Queries}, a query set anchored at the center of active voxels intersecting with the object surface, as shown in Fig.~\ref{fig:voxset}. Voxel Queries are not sampled on the surface but on a coarse voxel grid, so that their position can be easily obtained during test time by a coarse structure generation stage. Besides, the voxel center is decorrelated with different surfaces, reducing the training-test gap and greatly improving the generalization capabilities at test time.

\subsection{Detailed Geometry Generation}
\label{sec:method:gen}

As shown in Fig. \ref{fig:shape_arch}, we introduce a two-stage pipeline to fit the proposed VoxSet representation for generating geometry with ultimate details. The first stage generates the coarse sparse structure given the input image by voxelizing the results of off-the-shelf pretrained 3D generators~\cite{zhao2025hunyuan3d}. The second stage generates sparse voxel latents anchored at the voxel centers of the previous sparse structure. 

\textbf{Semi-Structured Geometry VAE.} We adopt the proposed VoxSet representation to train a semi-structured geometry VAE. As illustrated in Fig.~\ref{fig:voxset}, our method combines the strengths of VecSet~\cite{zhang20233dshape2vecset} and SLAT~\cite{xiang2024structured}, keeping latent tokens compact and structural. To support multi-resolution voxel structures, instead of randomly sampling voxel queries at various resolutions, we propose a simpler approach that supports arbitrary resolutions.
Specifically, we jitter the point queries during training by adding a small random offset $\epsilon \sim U\left[\frac{-1}{2R}, \frac{1}{2R}\right]$, where $R$ is the smallest resolution we aim to support. 
At test time or during diffusion training, voxel queries can be sampled at any resolution greater than $R$. 
The other aspects of our VAE are the same as in Hunyuan3D-2~\cite{zhao2025hunyuan3d}, except we only sample queries from a uniformly sampled point cloud.

\textbf{Adding Structure to Diffusion Transformer.} Following Hunyuan3D-2~\cite{zhao2025hunyuan3d}, we utilize a rectified-flow transformer to generate the VoxSet. Instead of solely conditioning on the input image, we propose to utilize the structure of VoxSet by adding rotary positional embedding (RoPE)~\cite{su2024roformer} to each noisy latent token. This change, despite being inconspicuous at first glance, is crucial in improving model convergence. The reason behind this can be two-fold, firstly, the amount of available 3D data is much smaller than 2D image and video counterparts, which makes the latent space severely unoccupied. Secondly, geometry generation is drastically different and a more difficult task than image generation due to its sparsity, \ie, the 3D geometric surface occupies only a small portion of its bounding box while every pixel in the image has an RGB value. As a result, previous approaches~\cite{zhao2025hunyuan3d,zhang2024clay} that only use a single image as condition could hardly guide the denoising trajectory towards detailed geometry. 
To reduce the training cost, we introduce two simple strategies: (1) instead of utilizing all structure tokens, we randomly sampled a fixed number of tokens during the training, which is much smaller than sparse voxel methods~\cite{xiang2024structured};
(2) we adopt a progressive training strategy by first training on 1024 tokens and progressively scaling up to 6144 tokens.

\textbf{Image Conditioning.}
Following Hunyuan3D-2~\cite{zhao2025hunyuan3d}, we use Dinov2-Giant~\cite{oquab2023dinov2} for image conditioning, taking the last hidden layer embedding without the class token. Different from Hunyuan3D-2’s 518 resolution, we use 1022 for finer details. The object is cropped via a binary mask while keeping the aspect ratio to reduce token length. No extra positional embedding is added, as Dino already encodes sufficient spatial information.

\begin{figure}[t] 
  \centering
\includegraphics[width=\linewidth]{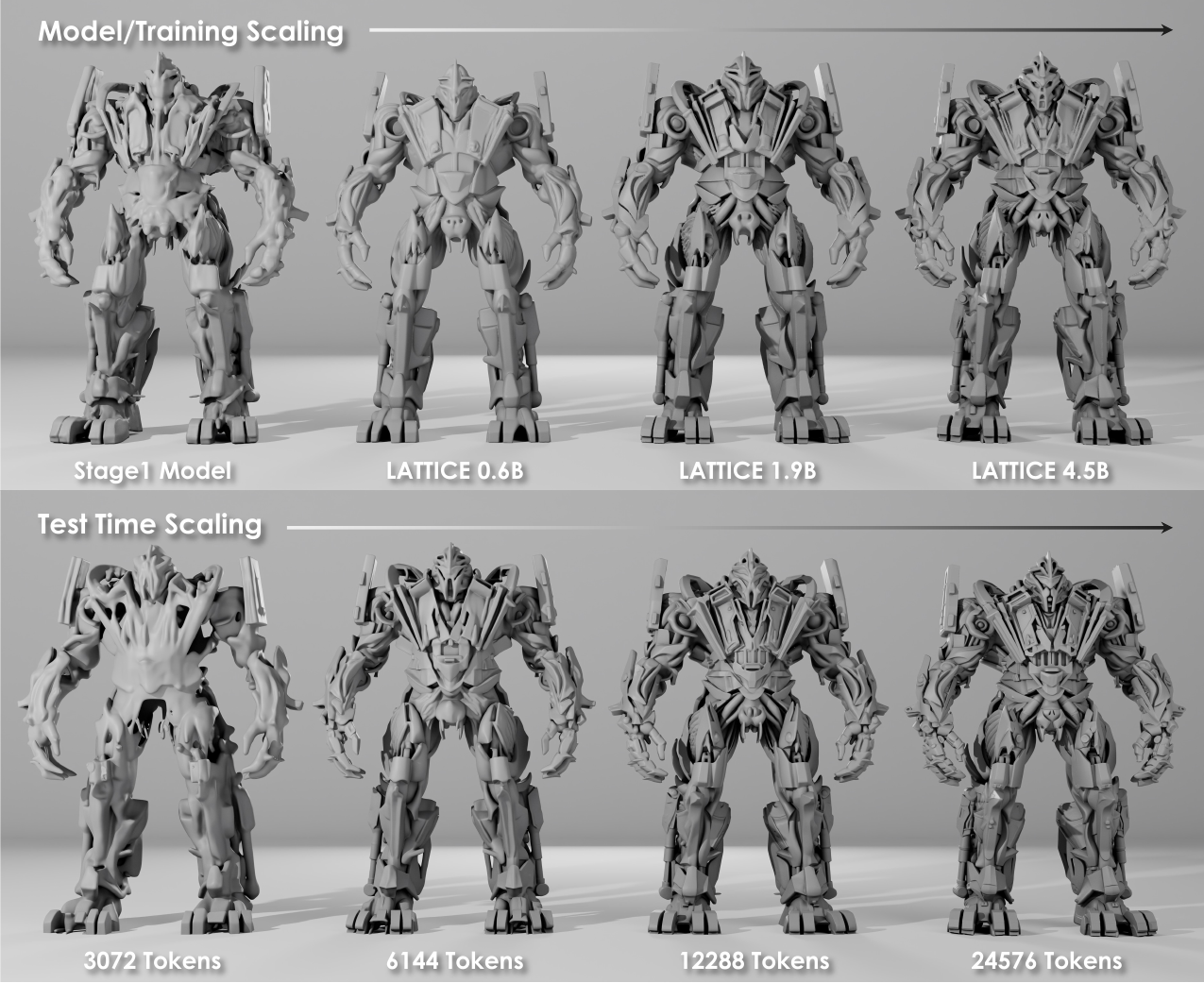}
\vspace{-3mm}
  \caption{Illustration of model/training and test scaling effects.} 
  \label{fig:scaling}
\end{figure}

\textbf{Training and Test-time Scaling.} We train several models in different model sizes, ranging from 0.6B to 4.5B. As shown in Fig.~\ref{fig:scaling}, our model exhibits stable scaling effect -- the bigger the better. 
Moreover, our model surprisingly reveals test-time scaling effect in token length as shown in Fig.~\ref{fig:scaling}. Even though our model is trained on 6144 tokens, we could increase the number of tokens to 12288, 24576, and even more by sampling more voxel queries.

\begin{figure}[t] 
  \centering
  \includegraphics[width=1\linewidth]{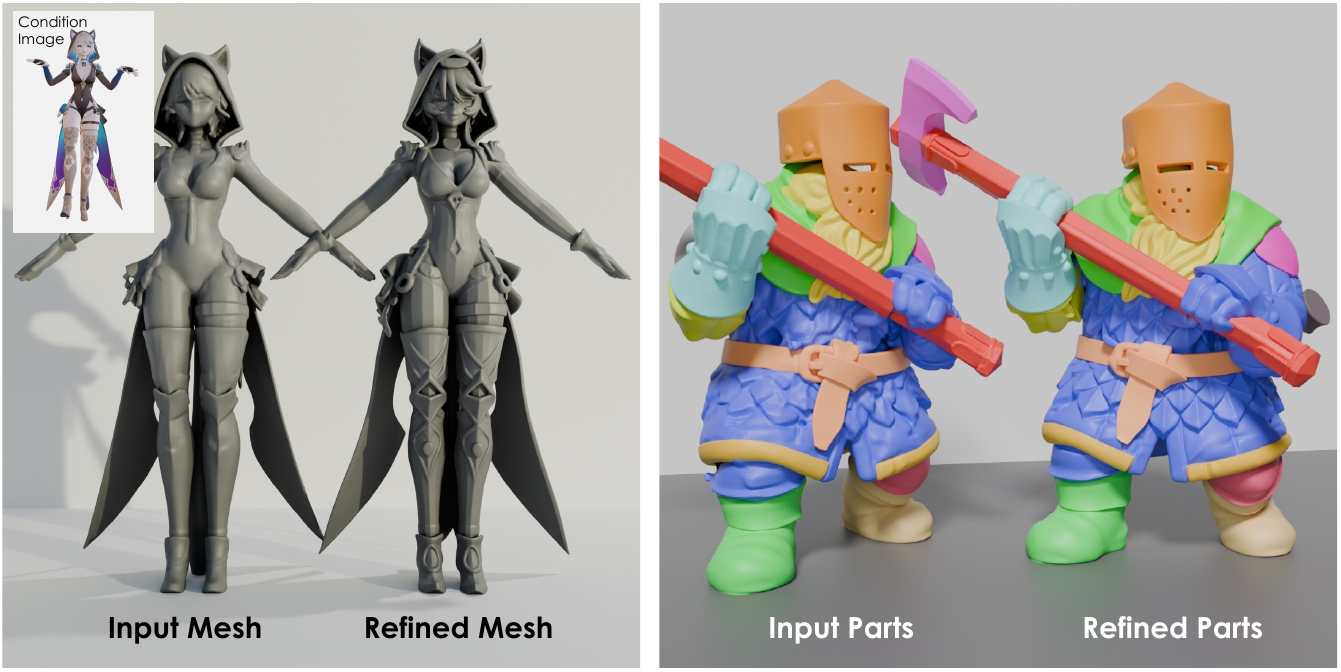}
  \caption{Illustrations of applications of \shortname. Mesh refinement in the left and part refinement in the right.} 
  \vspace{-3mm}
  \label{fig:application}
\end{figure}

\begin{figure*}[tbp] 
  \centering
\includegraphics[width=1\linewidth]{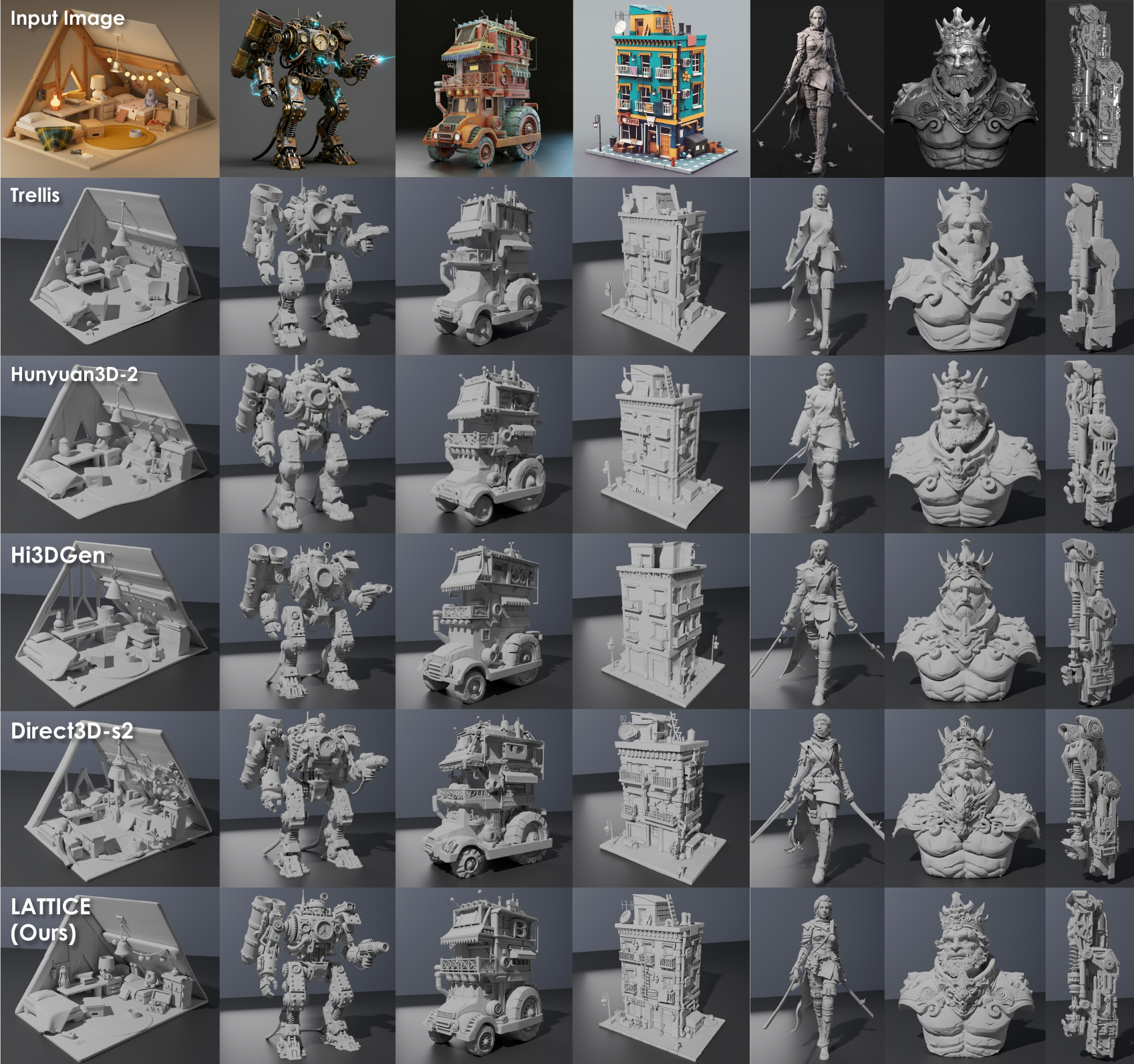}
  \caption{Visual comparison of geometry generation against several state-of-the-art open-source methods.} 
  \label{fig:cmp_geo_generation}
\end{figure*}

\subsection{Applications}

Thanks to the flexible design of the proposed architecture. 
We could adapt our model for various tasks.

\textbf{Mesh Refinement} can be extended to a broader context where the image is not aligned or missing, such as controllable generation or part refinement~\cite{yan2025x} as in Fig.\ref{fig:application}.    

\textbf{Mesh Editing} is also possible by manipulating the voxel queries and latent features of the given mesh, using the idea of Repaint~\cite{lugmayr2023inpainting}, MastControl~\cite{cao2023masactrl}, \etc.

\section{Experiments}

\subsection{Reconstruction}

We adopt two metrics for evaluating geometry reconstruction performance, including Chamfer Distance (CD) and F-score with a threshold of 0.001. To evaluate the reconstruction accurately, we use points-to-surface distances to calculate the metrics, with the mesh normalized to the range [-1, 1]. Similar to Dora~\cite{chen2024dora}, we construct a benchmark containing more challenging, detailed assets as LATTICE-Bench(R).
The competing methods consist of (1) representative VecSet-based methods: Hunyuan3D-2~\cite{zhao2025hunyuan3d}; and (2) Voxel-based methods: SparseFlex~\cite{he2025sparseflex}, and Direct3D-s2~\cite{wu2025direct3d}.
The numerical comparison is shown in Tab.~\ref{tab:geo_recon}, the metrics are multiplied by $10^{4}$ and $10^{2}$. Our method delivers top performance with a much more compact latent representation than voxel-based methods.

\begin{table}[t]
     \centering
     \small
     \setlength{\tabcolsep}{4pt}
     \begin{tabular}{r|cc|cc}
     \hline
     \textbf{Method} & \textbf{Res-} & \textbf{Latent Size} & \textbf{CD($\downarrow$)} & \textbf{F1($\uparrow$)} \\ 
     \hline
     Hunyuan3D-2~\cite{zhao2025hunyuan3d}      & N/A      & $64\times4096$       & 12.35 & 82.78 \\ 
                                               & N/A      & $64\times8192$       & 9.157 & 91.57 \\ 
     \hline
     SparseFlex~\cite{he2025sparseflex} &512              
                  & $8\times48557$       & 8.020 & 90.94 \\ 
        & 1024          & $8\times196028$       & 2.972 & 97.76 \\ 
     Direct3D-s2~\cite{wu2025direct3d}        & 1024       & $64\times46592$       & 4.987 & 97.46 \\ 
     \hline
     \textbf{LATTICE}~(Ours)& N/A & $64\times4096$  & 5.321 & {95.31} \\
                     & N/A  & $64\times8192$ & 2.909  & 98.53 \\
                     & N/A  & $64\times20480$ & \textbf{1.893} & \textbf{99.59} \\
     \hline
     \end{tabular}
     \vspace{-2mm}
     \caption{Quantitative comparisons of geometry reconstruction.}
     \label{tab:geo_recon}
\end{table}

\begin{table}
\centering
\small
\setlength{\tabcolsep}{5pt}
\begin{tabular}{r|cccc}
\hline
               \textbf{Method}         & \textbf{ULIP-T} & \textbf{ULIP-I} & \textbf{Uni-T} & \textbf{Uni-I}\\ \hline
Michelangelo~\citep{zhao2024michelangelo} & 0.075 & 0.115 & 0.213 & 0.261 \\
Craftsman 1.5~\citep{li2024craftsman}    & 0.074 & 0.129 & 0.237 & 0.298 \\
Trellis~\citep{xiang2024structured}      & 0.076 & 0.126 & 0.249 & 0.311 \\
Hunyuan3D 2.0~\citep{zhao2025hunyuan3d} & {0.077} & {0.130} & {0.251} & {0.315} \\ 
Hi3DGen~\cite{ye2025hi3dgen}  & 0.066 & 0.112  & 0.246 & 0.299 \\
Direct3D-s2~\cite{wu2025direct3d} & 0.074 & 0.122 & 0.247 & 0.314  \\
\hline
\textbf{LATTICE}-1.9B                  & \textbf{0.078} & \textbf{0.130}  & \textbf{0.254}   & \textbf{0.315}  \\ 
\hline
\end{tabular}
\vspace{-2mm}
\caption{Numerical comparison of geometry generation performance on ULIP~\cite{xue2023ulip} and Uni3D~\cite{zhou2023uni3d} similarities.}
\label{tab:geo_generation}
\end{table}

\subsection{Generation.} 
\begin{figure}[t] 
  \centering
\includegraphics[width=\linewidth]{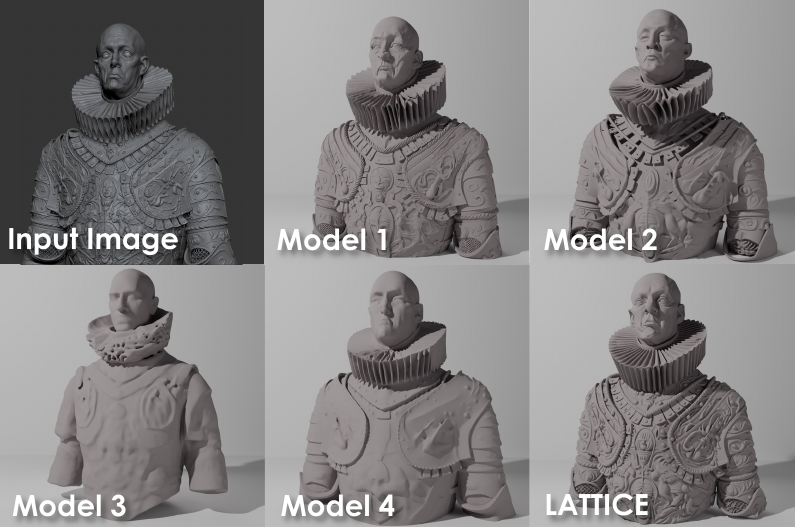}
  \caption{Visual comparison against commercial models.} 
  \label{fig:commercial}
\end{figure}

We evaluate the image-to-geometry generation through various metrics including ULIP~\cite{xue2023ulip}, Uni3D~\cite{zhou2023uni3d} for text-mesh and image-mesh similarities, following Hunyuan3D-2~\cite{zhao2025hunyuan3d}.
We compare our method against (1) open-source methods, Michelangelo~\cite{zhao2024michelangelo}, Craftman 1.5~\cite{li2024craftsman}, Trellis~\cite{xiang2024structured}, Hunyuan3D-2~\cite{zhao2025hunyuan3d}, Hi3DGen~\cite{ye2025hi3dgen}, and Direct3D-s2~\cite{wu2025direct3d};
(2) closed-source methods, which we denote as Model 1-4. 
The numerical comparison is shown in Tab.~\ref{tab:geo_generation}, omitting closed-source methods as obtaining a large amount of their results is very expensive. We compare \shortname-1.9B, which is the closest size to other models.
It can be observed that our method achieves the best performance.
Fig.~\ref{fig:cmp_geo_generation} and Fig.~\ref{fig:commercial} demonstrates the visual comparison, which confirms the superiority of our method.

\begin{figure}[t] 
  \centering
\includegraphics[width=\linewidth]{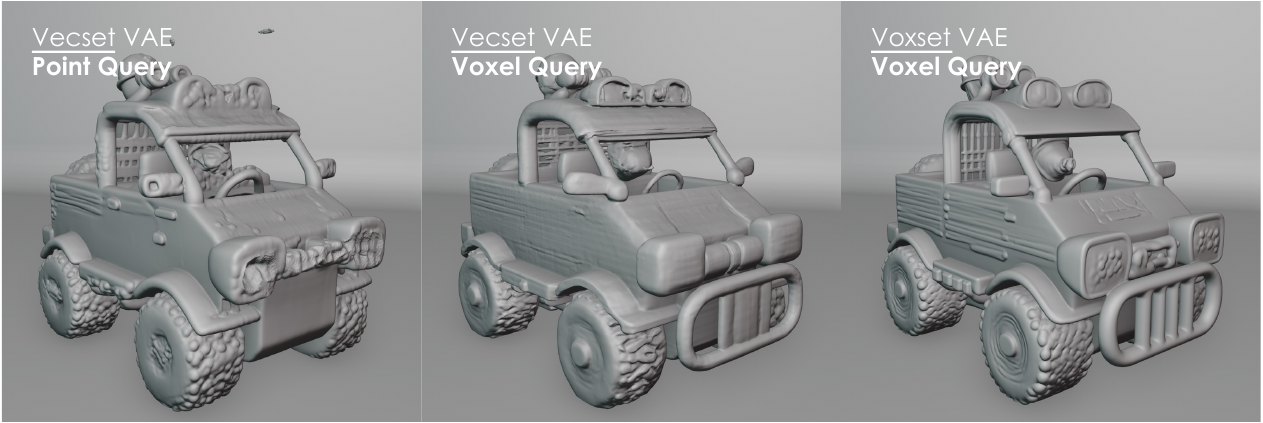}
  \caption{Ablation study on the proposed voxel query and VoxSet VAE, by incrementally adding each component.} 
  \label{fig:ablation_query}
\end{figure}

\subsection{Evaluation}

\begin{figure}[t] 
  \centering
  \includegraphics[width=\linewidth]{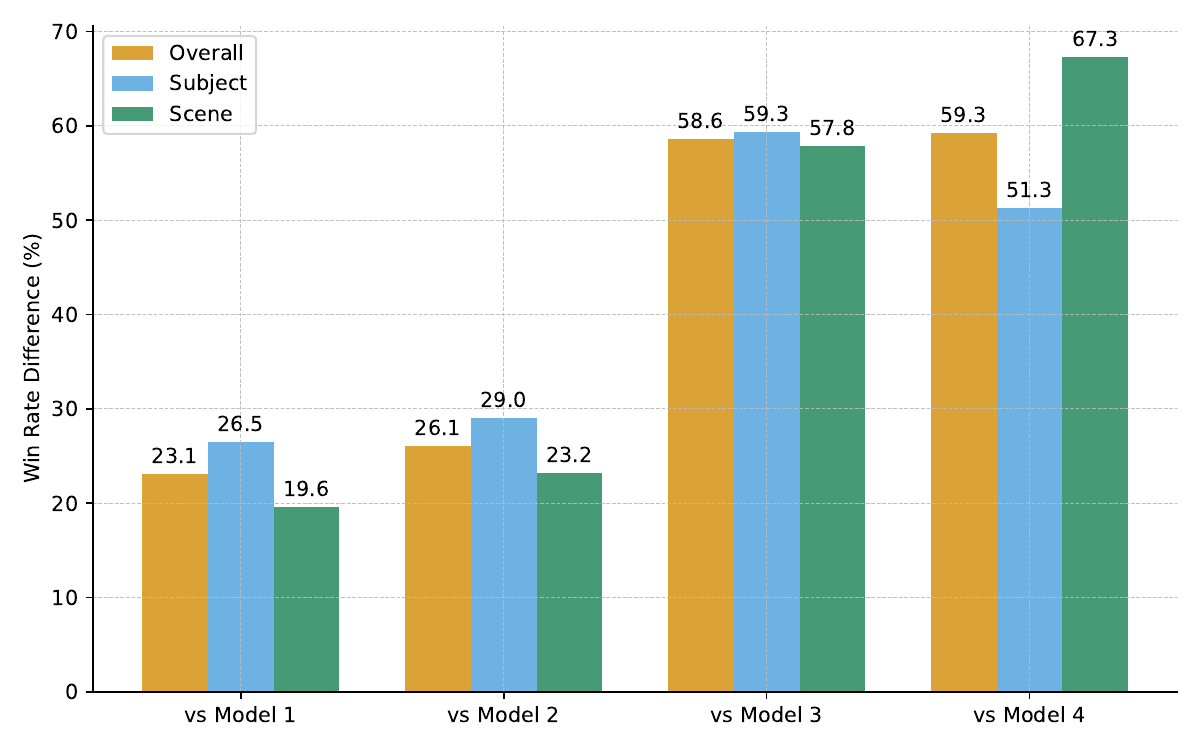}
  \caption{User study of our method against competitors showing win rate (\%) across Overall, Subject, and Scene categories.} 
  \label{fig:user_study}
\end{figure}
\begin{table}[]
\centering
\small
\setlength{\tabcolsep}{4.5pt}
\begin{tabular}{r|cc|ccc|ccc}
\hline
               & \multicolumn{2}{c|}{\textbf{Baseline}} & \multicolumn{3}{c|}{\textbf{+ Fixed Train}} & \multicolumn{3}{c}{\textbf{+ Query Jitter}} \\ \cline{2-9} 
\textbf{Res-}   & 64 & 128 & 64 & 128 & 256 & 64 & 128 & 256 \\ \hline
CD($\downarrow$) & 10.7 & 7.72 & 6.42 & 5.73 & 5.69 & 6.03 & 5.32 & 5.36  \\
F1($\uparrow$) & 85.3 & 91.4 & 92.9 & 94.5 & 94.7 & 93.7 & 95.3 & 95.3 \\
\hline
\end{tabular}
\caption{Ablation study of VAE training strategies. All settings are tested with 4096 tokens and voxel queries.}
\label{tab:vae_ablation}
\end{table}

\textbf{Effect of Voxel Queries.} To assess the effectiveness of the proposed voxel queries, we compare three DiTs, each utilizing different VAEs and query types, as shown in Fig.~\ref{fig:ablation_query}. All DiTs were trained for 200k steps (100k on 1024 tokens and 100k on 3072 tokens). The results indicate that voxel queries produce fewer artifacts, benefiting from a reduced domain gap, and the VoxSet VAE introduces more detail thanks to better reconstruction capability.

\textbf{Effect of Query Jitter.} Voxel queries are essential for bridging the gap between training and testing results in the first stage. To evaluate their impact, we ablate several VAEs with voxel queries by assessing their reconstruction performance. The numerical comparison is presented in Tab.~\ref{tab:vae_ablation}. As shown, the original point-query VAE suffers a significant degradation when tested with voxel queries. In contrast, the Query Jitter VAE outperforms VAEs trained at a fixed resolution and offers greater flexibility when applied to varying resolutions.

\textbf{User Study.} We also conducted a user study to assess human preferences across different methods. As shown in Fig.~\ref{fig:user_study}, our method was compared to four commercial models. The results clearly show that our method significantly outperforms the others.

\section{Conclusion}

We have presented \shortname, a novel framework that advances 3D asset generation by introducing Voxset, a semi-structured latent representation. By conditioning on localizable position information, we address key challenges in computational complexity, scalability, and fidelity for diffusion generation. Our method demonstrates superior performance in generating high-quality meshes, achieving stunning detail, smoothness, and sharpness. With its flexible encoding, low-cost training, and strong test-time scaling, \shortname represents a significant step forward in the automated generation of scalable, high-fidelity 3D content.

\newpage
\appendix

\section{Discussions}

\subsection{Scaling Behavior on Different Architectures}

We observe that 3D generation architectures differ substantially in their scaling behavior. Models without explicit localizable guidance (e.g., VecSet-based) scale much less effectively than those equipped with it, such as our proposed VoxSet architecture. Here, we provide a thorough analysis and accompanying demonstrations.

\begin{figure}[t] 
  \centering
\includegraphics[width=\linewidth]{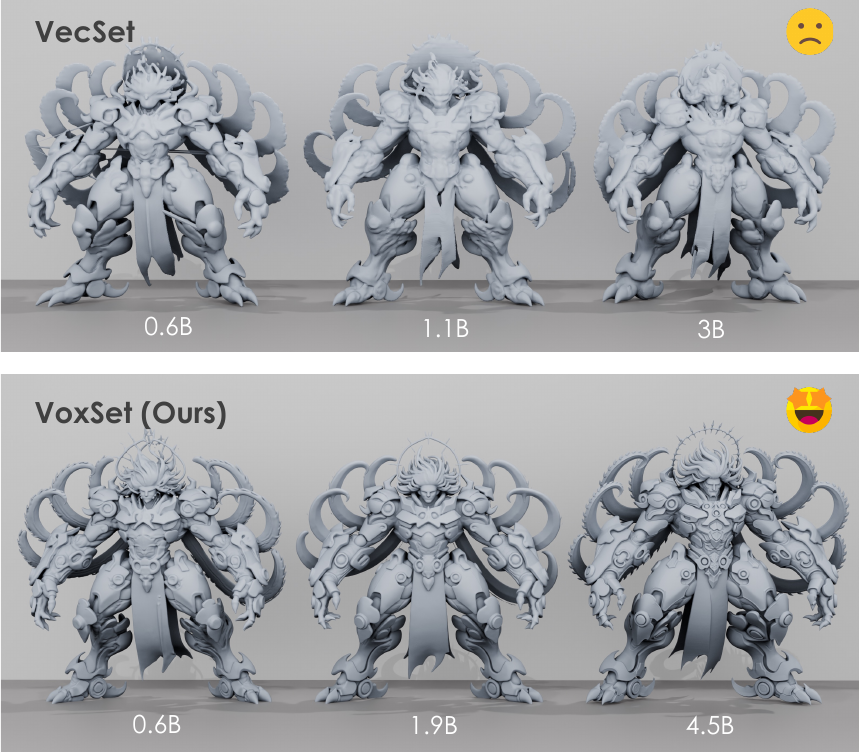}
  \caption{Illustration of the effect of model scaling (in parameters) on performance. VecSet models show limited improvement as parameters increase, whereas larger VoxSet models produce finer and more detailed results.} 
  \label{fig:model_scaling}
\end{figure}

\textbf{Model Scaling on Parameters.} 
Our first observation is that VecSet models hardly benefit from an increased number of parameters. To investigate this, we compare three popular VecSet models of different sizes: Hunyuan3D-2-mini~\cite{lai2025flashvdm} (0.6B), Hunyuan3D-2~\cite{zhao2025hunyuan3d} (1.1B), and Hunyuan3D-2.1~\cite{hunyuan3d2025hunyuan3d} (3B). The visual comparison is shown in Fig.~\ref{fig:model_scaling} (top row). Surprisingly, the three results are largely similar, with Hunyuan3D-2-mini (0.6B) even showing slightly better performance. This suggests that increasing model size may not play a decisive role for VecSet-based architectures.
In contrast, we find that the proposed VoxSet models consistently benefit from increased model size. To demonstrate this, we train three models of different sizes (0.6B, 1.9B, and 4.5B). The generated results are shown in Fig.~\ref{fig:model_scaling} (bottom row). It can be observed that as the number of parameters increases, the outputs exhibit more details, with sharper, smoother, and more regular structures.

Overall, these results demonstrate the effectiveness of the proposed VoxSet architecture, revealing that proper conditioning is a key factor in unlocking the benefits of model scaling. In other words, larger models are useful only when there is a clear correspondence between conditions and outputs; otherwise, increasing the number of parameters provides little advantage, as the model is more good at memorizing rather than abstracting and reasoning.
From the data perspective, the difference in scaling behaviors could also be explained by the lack of large-scale 3D data. We need more data to cover mappings with higher degrees of freedom, such as those with fewer conditions.

\begin{figure}[t] 
  \centering
\includegraphics[width=\linewidth]{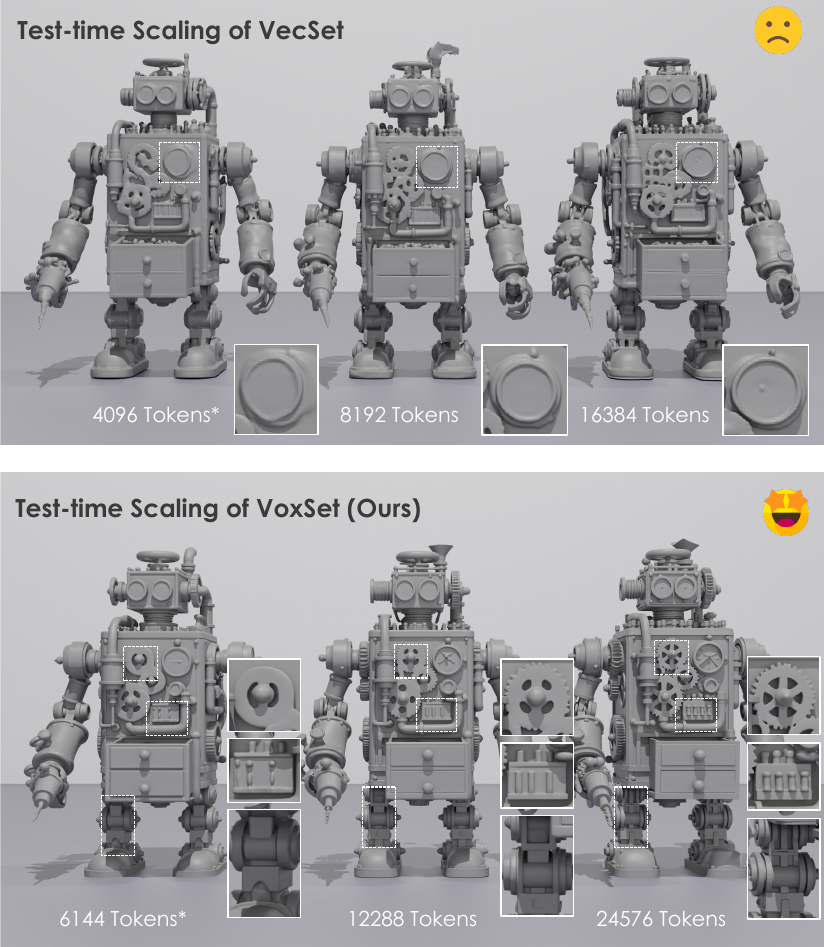}
  \caption{Illustration of the effect of test-time scaling (in shape tokens) on model performance. VecSet models exhibit limited gains as the number of tokens increases, showing early saturation. In contrast, VoxSet models consistently benefit from higher token counts, producing finer details and demonstrating stronger scaling capability. * indicates the token count used during training.} 
  \label{fig:test_scaling_cmp}
\end{figure}

\textbf{Test Time Scaling on Tokens.} 
Previously, we demonstrated that our model exhibits a strong test-time scaling effect on shape tokens in Sections 1 and 3.2. Specifically, the model trained with a maximum token length of N can be directly evaluated using 2N, 3N, or even more tokens during inference—no additional training or configuration is required.
Similar to model scaling, we here compare the test-time scaling behaviors of VecSet models and our VoxSet models, with results presented in Fig. \ref{fig:test_scaling_cmp}. The VecSet model was trained with 4096 tokens, while our VoxSet model was trained with 6144 tokens. We evaluated their generation performance at token lengths of 1N, 2N, and 4N.
As observed, the VecSet model benefits slightly from scaling tokens to 2N (e.g., improved round structures on the body), but further scaling yields negligible gains. In contrast, our VoxSet model consistently benefits from increased token counts: it shows improvements when scaling from N to 2N, and further gains when scaling from 2N to 4N. Notably, more tokens directly translate to richer details.

In general, the token scaling capability of both VecSet and VoxSet models originates from their ability to perform arbitrary-resolution autoencoding. During VAE training, we randomly sample queries across the entire 3D voxel grid or object surface—this process acts as a form of random dropout for full-token training. As a result, our VAE inadvertently acquires the ability to encode objects with any token length, even though this property was not an explicit design goal.
The same principle applies to DiT training: the random selection of queries also endows DiT with test-time scaling capability. Notably, VoxSet exhibits a more pronounced scaling effect, which can be attributed to its stronger correspondence between query locations and content. During training, the transformer learns to model the relationship between spatial positions and the tokens that should be generated at those positions.

\subsection{Representation Centric to Generation Centric}

In this paper, our model challenges two common takes, which is largely from the view of representation, including \emph{(1) Global vs Local}: ``VecSet is global that is better for overall shape and sparse voxel is local which is more suitable for details", and \emph{(2) Structural Sparse Grid vs Unstructural Set}: ``the sparse voxel is structural which is better for editing and other downstream tasks, VecSet is unstructural for these tasks."

\textbf{Localizable Code is All You Need.} 
It is evident that we can combine the strengths of both approaches through the proposed VoxSet—a semi-structured set-based representation that encodes global information. Nevertheless, we wish to emphasize that while these two common perspectives hold true to some extent, critical nuances remain.
From a generation standpoint, the key to better generation performance lies in localizability: specifically, strong guidance that is accessible during test time. Locality may offer advantages for compression efficiency and reconstruction quality, but it is not the primary factor for generation tasks.
Regarding structure, sparse voxels are inherently structural, a property that benefits many tasks. However, this does not mean VecSet lacks structure. In fact, VecSet inherently contains structural information. Its only limitation is that this structure cannot be identified during test time. Thus, we propose VoxSet to circumvent this issue.

\section{Implementation Details}

\textbf{Training Setup.}
To evaluate the scaling effect of the proposed architecture, we train several models of various sizes, including a medium model (0.6B parameters), an XL model (1.9B), and an XXL model (4.5B). Unlike CLAY~\cite{zhang2024clay}, we do not adopt progressive model scaling; instead, all models are trained from scratch. Instead, we employ a multi-stage token scaling strategy. Within each stage, we use a constant learning rate with a linear warm-up, while gradually decreasing the base learning rate across stages from $1\times10^{-4}$ to $1\times10^{-6}$. The batch size is maximized to fit GPU memory, reaching up to 2048 in our experiments. We utilize ZeRO-based optimizer, gradient, and parameter partitioning from DeepSpeed~\cite{rajbhandari2020zero} to efficiently train large models on a distributed GPU cluster.
All models are trained using the flow matching objective with a linear coupling plan, following the formulation in SiT~\cite{ma2024sit}. 
Additionally, to enable classifier-free guidance~\cite{ho2022classifier}, we randomly replace conditioning embeddings with zero embeddings at a probability of 10\% during training.

\textbf{Data Preparation.}
Our data processing pipeline mainly includes three steps, (1) data filtering; (2) watertighting; (3) point-cloud sampling and SDF extraction. We apply extensive data filtering to improve the quality of the dataset, which includes removing AI-generated assets, scanned assets, extreme complex scenes, and assets with plane. We randomly sample millions of point cloud during the training, and split them into chunks to accelerate data loading.

\section{Post-Training}

\textbf{High-Quality Finetuning.} We introduce an additional finetuning stage with very high-quality data, which helps in improving details generation. All parameters in DiT are updated with a small learning rate. We filter the high-quality data by a combination of criterions including the number of faces, the number of sharp edges, and reconstruction quality, resulting in roughly 15k samples. 

\textbf{Model Acceleration.} We adopt FlashVDM~\cite{lai2025flashvdm} to accelerate the geometry VAE decoding. For diffusion sampling, benifit from strong structure guidance from RoPE, we find that our models are inherent few-step generators. Nevertheless, we still perform guidance distillation and step distillation to further reduce the sampling cost.

\section{User Study Setting}

To evaluate the perceptual quality of the generated results, we conducted a large-scale user study consisting of approximately 500 questions. For each question, three independent participants were asked to rank all the presented results according to their visual quality or fidelity. Each question displayed outputs from all compared methods in a randomized order to ensure fairness. Finally, we aggregated the rankings across all participants and questions to compute the winning rate of each method, which reflects its overall preference by human evaluators.

\begin{figure*}[t] 
  \centering
\includegraphics[width=\linewidth]{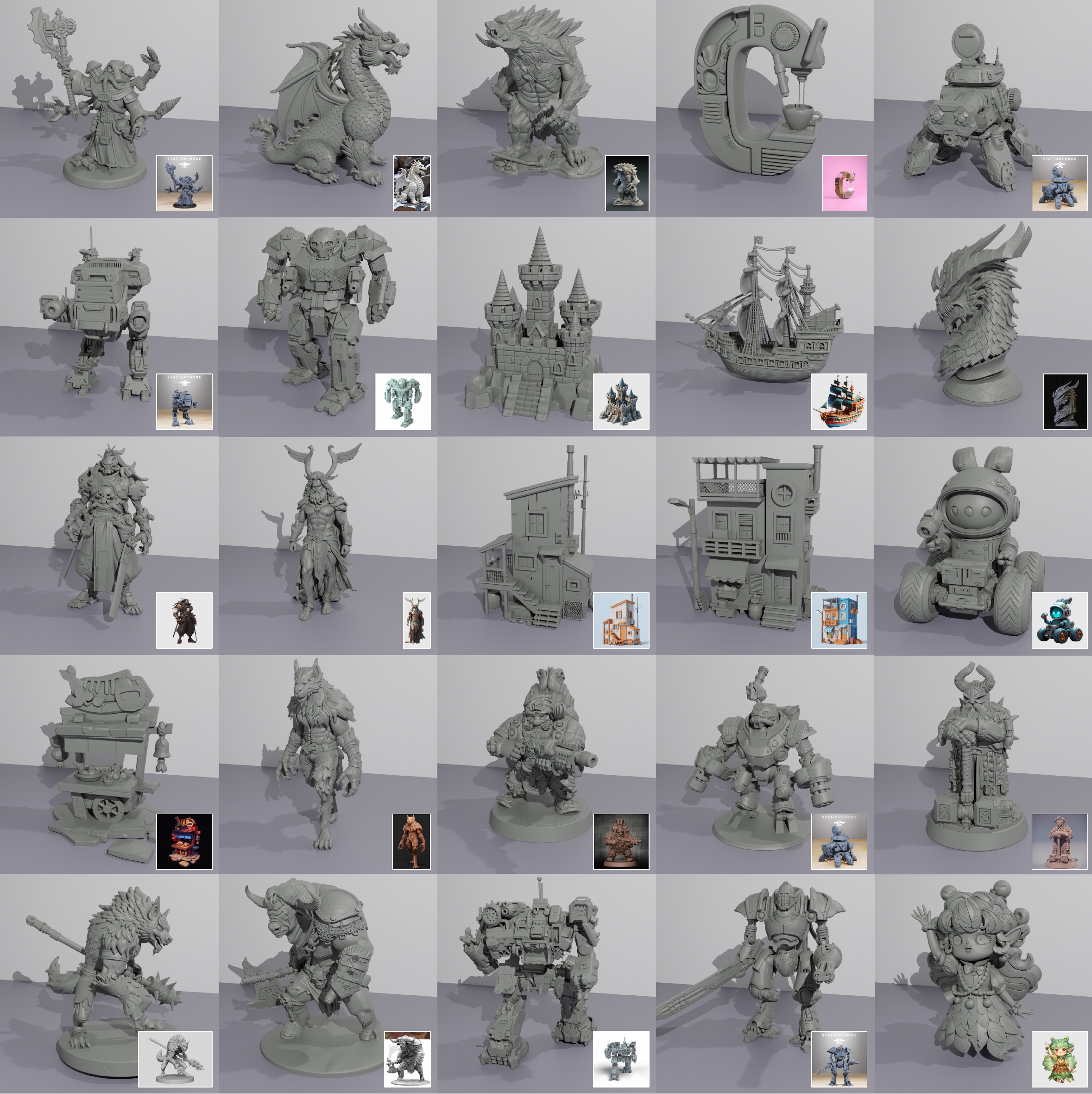}
  \caption{More visual results for image-to-geometry generation of \shortname.} 
  \label{fig:more_shape_results}
\end{figure*}

\section{More Results}

Fig.~\ref{fig:more_shape_results} shows additional results of our model. All examples are presented without cherry-picking.

{
    \small
    \bibliographystyle{ieeenat_fullname}
    \bibliography{sample-base}
}

\end{document}